\documentclass{aa}

\usepackage{amssymb}
\usepackage[usenames,dvipsnames]{xcolor}
\usepackage{graphicx}
\usepackage{nicefrac}
\usepackage[rightcaption]{sidecap}
\usepackage{natbib}
\usepackage{txfonts}
\usepackage{url}
\usepackage{hyperref}
\hypersetup{
    final=true,
    pageanchor=true,
    colorlinks=true,
    breaklinks=true,
    linkcolor=blue,
    citecolor=blue,
    urlcolor=blue,
    pdfpagemode=UseNone,
    pdftitle={X-Class Flares NOAA 21673},
    pdfauthor={M. Verma},
    pdfsubject={Solar Physics},
    pdfkeywords={Sun: chromosphere, Sun: photosphere, Sun: surface magnetism,
    Sun: sunspots, Techniques: image processing, Methods: data analysis}}

\begin{document}


\title{On the origin of two X-class flares in active region NOAA~12673}
\subtitle{Shear flows and head-on collision of new and pre-existing flux}

\author{M.\ Verma}

\institute{%
    Leibniz-Institut f{\"u}r Astrophysik Potsdam (AIP),
    An der Sternwarte~16,
    14482 Potsdam,
    Germany\\
    e-mail: \href{mailto:mverma@aip.de}{\texttt{mverma@aip.de}}}

\authorrunning{M.\ Verma}

\date{Received 31/10/2017; accepted 24/01/2018}

\abstract
{Flare-prolific active region NOAA~12673 produced consecutive X2.2 and X9.3 flares on 6~September 2017. To scrutinize the morphological, magnetic, and horizontal flow properties associated with these flares, a 7-hour time-series was used consisting of continuum images, line-of-sight/vector magnetograms, and 1600~\AA\ UV images. These data were acquired with the Helioseismic and Magnetic Imager (HMI) and the Atmospheric Imaging Assembly (AIA). The white-light flare emission differed for both flares, while the X2.2 flare displayed localized, confined flare kernels, the X9.3 flare exhibited a two-ribbon structure. In contrast, the excess UV emission exhibited a similar structure for both flares, but with larger areal extent for the X9.3 flare. These two flares represented a scenario, where the first confined flare acted as precursor, setting up the stage for the more extended flare. Difference maps for continuum and magnetograms revealed locations of significant changes, i.e., penumbral decay and umbral strengthening. The curved magnetic polarity inversion line in the $\delta$-spot was the fulcrum of most changes. Horizontal proper motions were computed using the differential affine velocity estimator for vector magnetograms (DAVE4VM). Persistent flow features included (1) strong shear flows along the polarity inversion line, where the negative, parasitic polarity tried to bypass the majority, positive-polarity part of the $\delta$-spot in the north, (2) a group of positive-polarity spots, which moved around the $\delta$-spot in the south, moving away from the $\delta$-spot with significant horizontal flow speeds, and (3) intense moat flows partially surrounding the penumbra of several sunspots, which became weaker in regions with penumbral decay. The enhanced flare activity has its origin in the head-on collision of newly emerging flux with an already existing regular, $\alpha$-spot. Umbral cores of emerging bipoles were incorporated in its penumbra, creating a $\delta$-configuration with an extended polarity inversion line, as the parasitic umbral cores were stretched while circumventing the majority polarity.} 

\keywords{Sun: activity --
    Sun: flares --
    Sun: sunspots --
    Sun: magnetic fields}

\maketitle


\section{Introduction}

The relationship between sunspot numbers and flare occurrence as a function of 
the solar cycle was extensively examined, revealing a surge of renewed solar 
activity occurs a few years after the cycle 
maximum \citep{Fritzova1973, Svestka1995}. \citet{Temmer2003} noticed, in a 
correlation analysis of flare occurrence and sunspot number, that the flare 
occurrence lags behind the sunspot number. For solar cycle 23, \citet{Bai2006} 
found high flare activity in the late declining phase of the cycle. Not only an 
increased flare activity was noted but many low-latitude coronal holes were
repeatedly observed during the declining and minimum phases of solar cycle 
23 \citep{Abramenko2010}. In addition, \citet{Gopalswamy2003b} determined that the 
occurrence rate of coronal mass ejections (CMEs) peaked two years after the cycle 
maximum, and in the analysis of \citet{Kilcik2011}, the maximum speeds of CMEs 
also peaked approximately at that time. Recently, 
\citet{Lee2016} reported that the occurrence rates of major flares and  
front-side halo CMEs are higher during the descending phase of solar cycle 23 as 
compared to other phases of the cycle. The recent rejuvenation in the Sun's 
large-scale magnetic field during solar cycle 24 starting in the second half of 
2014, was pointed out by \citet{Sheeley2015}. They concluded that this was the
result of systematic flux emergence in active regions, whose longitudinal
distribution greatly increased the Sun's dipole moment.

The role of horizontal shear flows in building up magnetic stress in 
flare-productive active regions was already noticed by \citet{Harvey1976}. 
Horizontal shear flows are not only responsible for magnetic stress but are also 
drivers for magnetic reconnection \citep{Yurchyshyn2006}. \citet{Yang2004} 
measured the horizontal flows in active region NOAA 10486 before an X10 flare and 
found shear flows with local correlation tracking \citep[LCT,][]{November1988} 
of up to 1.6~km~s$^{-1}$ along the polarity inversion line (PIL). 
These strong shear flows were correlated with white-light flare kernels in the 
optical range. Strong horizontal shear flows, which decrease with height in the
atmosphere, were also observed by \citet{Deng2006} in the same active region. 
Based on photospheric flows and the flux around the PIL, \citet{Welsch2009} 
derived indicators for the occurrence of flares. Many case studies established 
crucial links between photospheric shear flows, build-up of free-energy in a stressed
magnetic field topology, and major flares \citep[e.g.,][]{Tan2009, Liu2010}. 
More recently, \citet{Beauregard2012} and \citet{Wang2014} analyzed the horizontal 
flows in active region NOAA~11158 in relation to an X2.2 flare. The photospheric
signatures produced by this major flare were rather diminutive. On the other 
hand, during the impulsive phase shear flows around the PIL showed a sudden 
decrease and a change in the intrinsic rotation of the two main sunspots. 
Recently, \citet{Wang2015} reviewed shear motions, sunspot rotation, and 
flux emergence regarding their role in triggering solar eruptive events.  

\begin{figure}[t]
\centering
\includegraphics[width=\columnwidth]{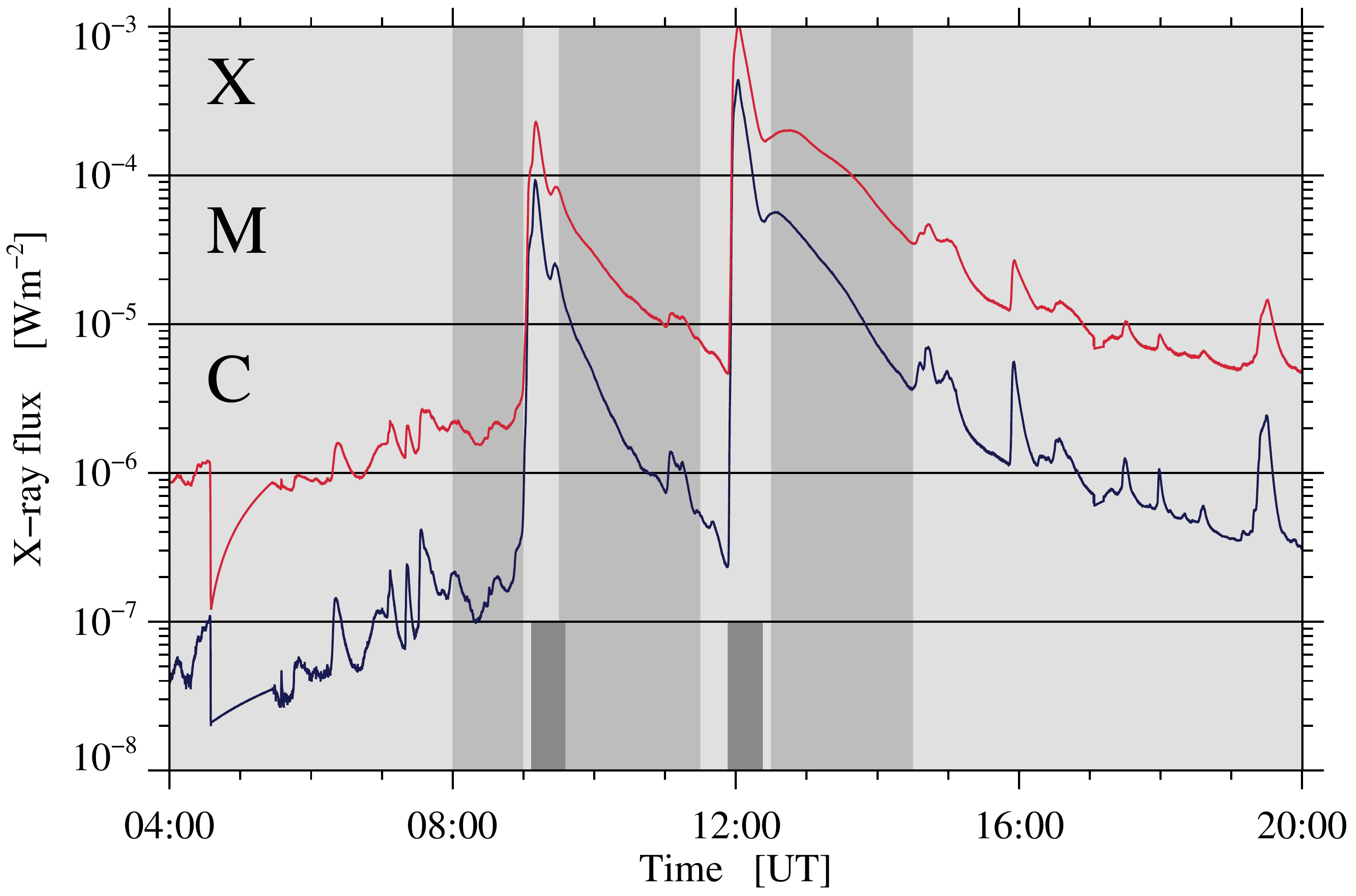}
\caption{GOES 13 X-ray flux on 6~September 2017 in the 0.1--\,0.8~nm 
    (\textit{top}) and 0.05\,--\,0.4~nm (\textit{bottom}) energy channels. 
    The three light gray regions indicate the pre-flare, interim, and 
    post-flare phases, which were used to compute horizontal proper motions 
    (Fig.~\ref{FIG04}). The two narrow, dark gray regions refer to the 
    time intervals for computing the white-light flare kernels 
    (Fig.~\ref{FIG02}).}
\label{FIG01}
\end{figure} 


\section{Observations and data reduction}

Active region NOAA~12673 appeared on the South-East limb of the solar disk on 
29~August 2017 as a simple $\alpha$-spot. Over the next few days, it maintained
its single-spot configuration. On 3~September 2017, new flux emerged as bipolar
regions following the $\alpha$-spot, and by the next day, the region had acquired
a complex $\beta\gamma$-configuration, when the positive polarity of the bipoles
merged with the main spot. The complexity of the region increased, and a 
$\delta$-configuration appeared by 5~September 2017, when the negative polarity of 
the bipoles collided head-on with the main spot. A unique aspect of this
flare-prolific active region was the short period of just three hours between
two major flares. The X2.2 flare occurred at 
08:57~UT followed by an X9.3 flare at 11:53~UT on 6 September (Fig.~\ref{FIG01}).
The latter was the strongest flare in solar cycle 24. A detailed 
description of the magnetic evolution is given by \citet{Yang2017}, \citet{Sun2017}
call attention to an overall extremely high flux emergence rate and non-potential 
magnetic field topology, and \citet{Wang2018} report extremely strong transverse
magnetic fields near the PIL of more than 5500~G.

The present work is based on HMI continuum images and magnetograms 
\citep{Scherrer2012} and UV images obtained by AIA \citep{Lemen2012},
both on board the Solar Dynamics Observatory \citep[SDO,][]{Pesnell2012}. 
These time-series cover a 7-hour period from  08:00\,--15:00~UT. The data were 
divided in three sets: ``pre-flare'' one hour before the X2.2 flare (08:00\,--\,09:00~UT),
``interim'' two hours between the two X-class flares (9:30\,--11:30~UT), and ``post-flare'' 
two hours after the X9.3 flare (12:00\,--14:00~UT). The goal was to capture differences 
in the flow fields before, in between, and after the two major flares. The HMI and AIA 
data had a cadence of 45~s and 48~s, respectively, and the former were adapted to the AIA
image scale of 0.6\arcsec~pixel$^{-1}$. Images and magnetograms were then compensated for
differential rotation and rotated to the position, when the active region crossed the 
central meridian at 19:00~UT on 3~September 2017. The continuum and UV images were 
corrected for the center-to-limb variation and divided by a 2D limb-darkening function 
based on the average quiet-Sun intensity. 

\begin{figure}[t]
\includegraphics[width=\columnwidth]{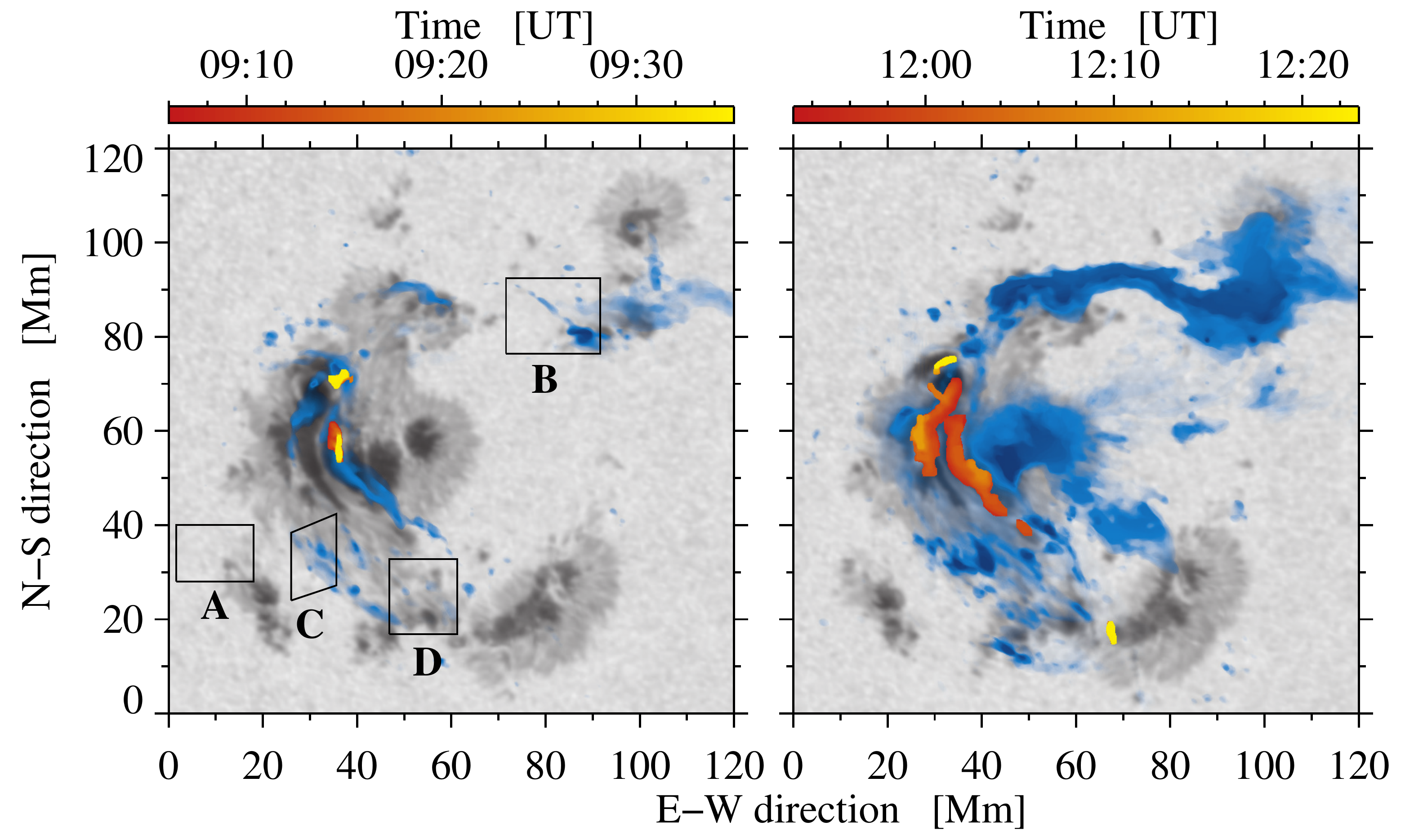}
\caption{Continuum images at 08:57~UT (\textit{left}) and 11:53~UT 
    (\textit{right}) superimposed with the white-light flare kernels derived 
    from 30-minute time-series of continuum images covering the X2.2 and 
    X9.3 flares (see scale bar for the temporal evolution). The
    blue-colored overlays represent the 2D frequency distribution of pixels
    exhibiting strong UV excess brightenings with respect to the pre-flare
    phase. The boxes mark the locations of penumbral decay
    (\textsf{A\,--\,C}) and umbral strengthening (\textsf{D}).}
\label{FIG02}
\end{figure} 

For tracking the plasma flows we used the Space-weather HMI Active Region
Patches (SHARP) vector magnetogram data product with 12-minute cadence 
\citep{Bobra2014}. The SHARP maps were generated from polarization measurements
at six wavelengths along the Fe\,\textsc{i} 6173~\AA\ spectral line \citep{Hoeksema2014}. 
The data were inverted using the Very Fast Inversion of the Stokes Vector (VFISV) 
algorithm \citep{Borrero2011a} based on the Milne-Eddington approximation, and the 
180$^\circ$ ambiguity in azimuth was resolved using the minimum-energy code 
\citep{Metcalf1994, Leka2009}. The inversion provided several physical parameters, 
including maps of continuum intensity, magnetic field, azimuth, inclination, and 
line-of-sight (LOS) velocity (see online movie). Since only six wavelength points 
are used in the inversion of HMI 
vector magnetograms the errors are usually large in the derived magnetic 
field parameters. In addition, an upper limit of 5000~G is set for the magnetic field
strength. The total magnetic field strength and transverse magnetic field in all 36 maps 
combined exceed 4000~G
only for about 2100 and 250 pixels, respectively. Thus, extremely strong magnetic fields, 
as reported by \citet{Wang2018}, were not observed. From the two types of available 
SHARP data products \citep{Bobra2014}, definitive maps were only available for the
interim and post-flare phases because of a semiannual spacecraft eclipse period. 
Thus, five near-real-time maps were used for the pre-flare phase. 
 
In the next step, images and magnetic field maps were corrected for 
geometrical foreshortening \citep{Verma2011}, which resulted in maps, where one 
pixel corresponds to 400~km $\times$ 400~km. Horizontal proper motions were 
derived with DAVE4VM \citep{Schuck2008} using convolution with the Scharr operator
and a five-point-stencil with a time difference of 12~min between neighboring 
magnetograms for the spatial and temporal derivatives of the magnetic field, 
respectively. The horizontal velocities were averaged over one hour for the 
pre-flare period, and two hours for the interim and post-flare periods.

\begin{figure*}[t]
\includegraphics[width=\textwidth]{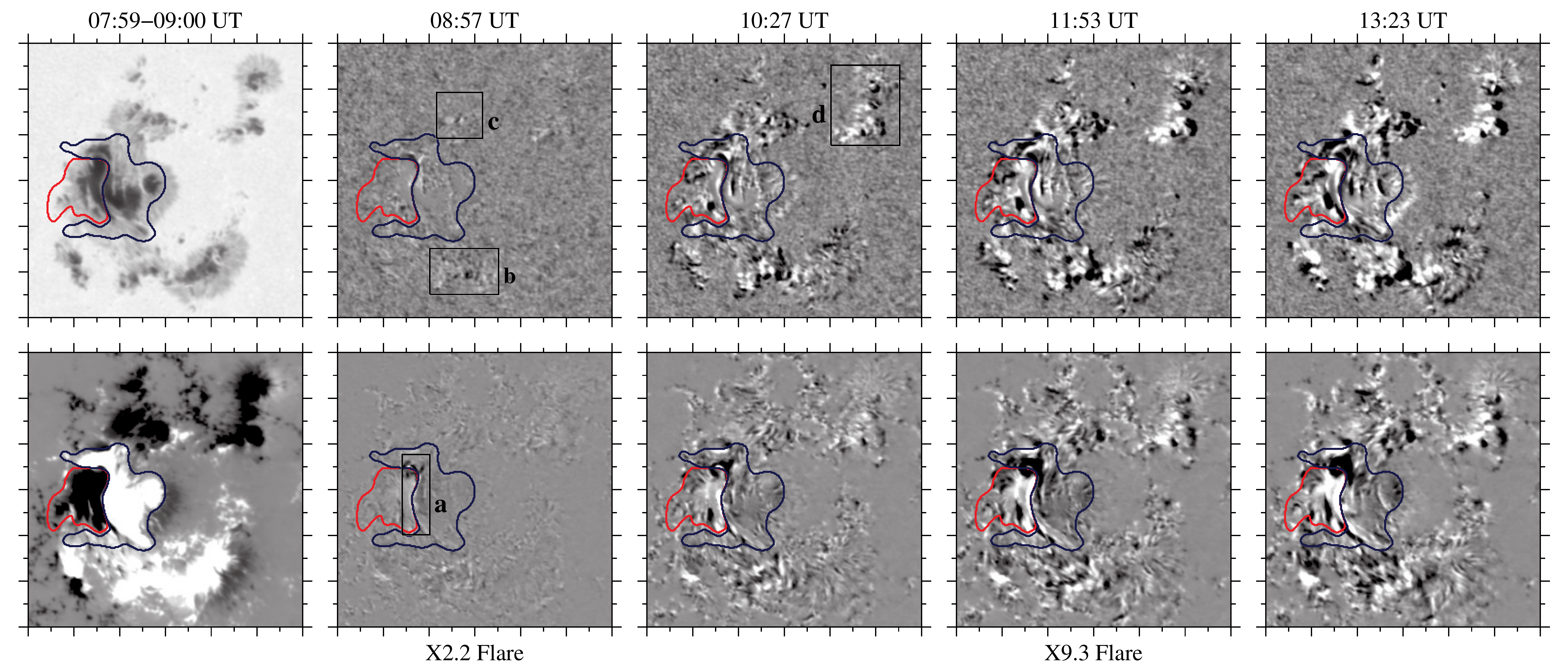}
\caption{One-hour averaged continuum image (\textit{top-left}) and
    magnetogram (\textit{bottom-left}) of active region NOAA~12673. The next 
    four panels (\textit{top and bottom}) covering an 
    area of 120~Mm $\times$
    120~Mm show difference maps scaled between $\pm 0.2$~$I_0$ (continuum 
    images) and $\pm 500$~G (magnetograms). The red and blue contours mark the
    boundaries of negative and positive polarities comprising the 
    $\delta$-spot. The intersection of these two colored lines demarcates the 
    PIL. The boxes labeled `a', `b', `c', and `d' mark locations of persistent
    changes.}
\label{FIG03}
\end{figure*} 


\section{Results}

Active region NOAA~12673 produced many C-, 27 M-, and four X-class flares 
in one week (4\,--10~September 2017). The two X-class flares on 6~September 2017 were
separated in time by only about three hours (Fig.~\ref{FIG01}). The immediate
succession of two major flares motivated this case study. The flare 
kernels for both flares were identified in continuum and UV images. 
To extract the photospheric continuum emission, average maps were created
using 40 continuum images before 09:06~UT and 11:53~UT for the X2.2 and X9.3
flares, respectively. These background maps were subtracted from individual
images covering the next 30~min starting at the flare initiation time. Choosing
a threshold of $+0.1\,I_{0}$, where $I_0$ is the quiet-Sun intensity, for the 
excess intensity was sufficient to unveil the continuum flare emission and its 
location. To obtain the areal extent with sub-pixel accuracy, the images were 
magnified by a factor of 2.5 using Fourier interpolation. Small-scale features 
were discarded and smooth boundaries were ensured using morphological opening. 
The temporal evolution of regions with continuum flare emission is given by 
the scale bars in Fig.~\ref{FIG02}, where red shows the beginning and yellow 
the end of the white-light flare.

An average map based on 50 images taken between 07:30\,--08:10~UT was used to
compute the excess UV emission after both flares. This background map was 
then subtracted from all individual images over a two-hour period in both the
interim and the post-flare phases. The flare ribbons and kernels are
characterized by very strong UV emission. Thus, a threshold of 3.5 times the
background emission was used to label the flaring regions. A 2D frequency
distribution of areas displaying significant UV excess brightenings was
superimposed in blue colors on the gray-scale continuum image in both panels 
of Fig.~\ref{FIG02}. Therefore, a strong saturation of the color blue 
indicates regions, which exhibit the strongest UV flare emissions.

During the X2.2 flare, the white-light flare kernels were confined and 
limited to two small patches near the PIL in the positive polarity of the
$\delta$-spot. These kernels appeared almost 10-min after the 
flare peak time. In contrast, the white-light kernels for the X9.3 flare
appeared 3~min after the X-ray emission peaked. They were more extended and 
formed a two-ribbon configuration on both sides 
of the PIL. These two ribbons migrated away from the PIL as indicated by the
color gradient from dark red to light orange in the right panel of
Fig.~\ref{FIG02}. Two remote brightenings (yellow) appeared at the periphery of 
two umbral cores away from the PIL towards the end of the 30-minute interval.
The excess UV emission for the X2.2 flare was mainly limited to the 
$\delta$-spot and was sparsely seen in neighboring sunspots. However, for 
the X9.3 flare, the flaring area was much larger, and the excess emission
stretched to all sunspots in the active region. Even though both X-class flares 
can be categorized as homologous flares, significant differences in their 
morphology and temporal evolution are apparent, both for the white-light 
flare kernels and the excess UV brightenings. The X2.2 flare likely
weakened the magnetic topology facilitating the next eruptive flare. Following
the evolution in time-lapse movies of the Large-Angle Spectrometric 
Coronagraph \citep[LASCO,][]{Brueckner1995}, a CME erupted only for the X9.3 
flare, whereas the X2.2 flare stayed confined.

\begin{figure*}
\centering
\includegraphics[width=\textwidth]{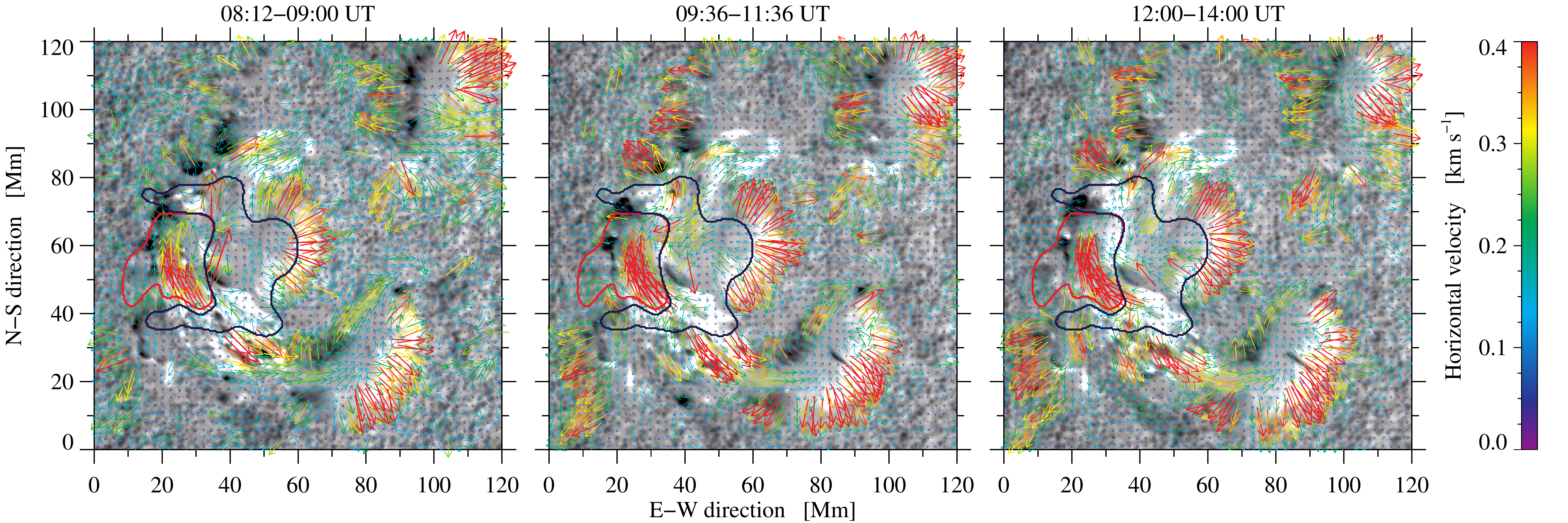}
\caption{Time-averaged horizontal proper motions in active region 
    NOAA~12673 computed with DAVE4VM for the pre-flare, interim, and 
    post-flare phases (\textit{left-to-right}). Color-coded vectors 
    indicate magnitude and direction of the horizontal flows. The 
    background images are LOS velocity maps based on spectral inversions at 
    08:36~UT, 10:36~UT, and 11:00~UT scaled between $\pm 1.4$~km~s$^{-1}$ 
    (black/white corresponds to blue/redshifts).}
\label{FIG04}
\end{figure*} 

Active region NOAA~12673 contained four major sunspots. The large, central
$\delta$-spot was the fulcrum of all flare activity. This spot differed from 
typical $\delta$-spots and was formed by the head-on collision of newly emerging 
bipolar regions with pre-existing flux in the form of a simple $\alpha$-spot. This 
created a complex active region \citep{Yang2017} with the main spot containing the 
$\delta$-configuration and three neighboring spots, i.e., two on top 
(`c' and `d' in Fig.~\ref{FIG02}) and spot `b' at the bottom, which also exhibited
peculiar proper motions. Within the $\delta$-spot a curved light-bridge `a' 
coincides with the PIL. The umbral cores on both sides of 
the light-bridge are elongated and curved, which already provides a first
indication of shear flows and twist within the flux system. This 
magnetically ``stressed'' configuration motivated this study, and difference
maps were computed for both continuum images and magnetograms 
(Fig.~\ref{FIG03}) to clearly identify, where the major changes occurred 
during and after the major flares.

The difference maps were computed by subtracting time-averaged maps from
individual continuum images and magnetograms. Before the X2.2 flare only 
minute changes occurred in both continuum intensity and magnetic field, with 
some strengthening of the magnetic field in the upper part of the PIL. 
Starting with the X2.2 flare, these changes became more apparent in this
location, where the reduced/strengthened magnetic fields in positive/negative 
polarities across the PIL appeared as an \textsf{X}-shaped feature. One and a half
hours after the X2.2 flare, considerable changes took place in the vicinity of
the PIL. A small dark localized region at coordinates (25~Mm, 50~Mm) 
appeared in both types of difference maps in the negative polarity of the 
$\delta$-spot. This is caused by the parasitic polarity trying to bypass the 
dominant positive polarity of the $\delta$-spot, which leads to flux cancellation.
The top of the PIL retained the reduced magnetic field. At this time, prominent 
changes in intensity and magnetic field became evident in the neighboring sunspots 
`b', `c', and `d' as well. In the positive-polarity spot `b' and 
negative-polarity spots `c' and `d', the decay of the penumbra on the side 
facing the $\delta$-spot was evident in both difference maps. Umbral strengthening
in the immediate vicinity of penumbral decay was seen in spots `b' and `d'. 
During and after the X9.3 flare, the changes in all marked locations remained, which 
indicates that the magnetic topology of the active region reached a new
equilibrium.

\begin{SCfigure*}
\includegraphics[width=0.74\columnwidth]{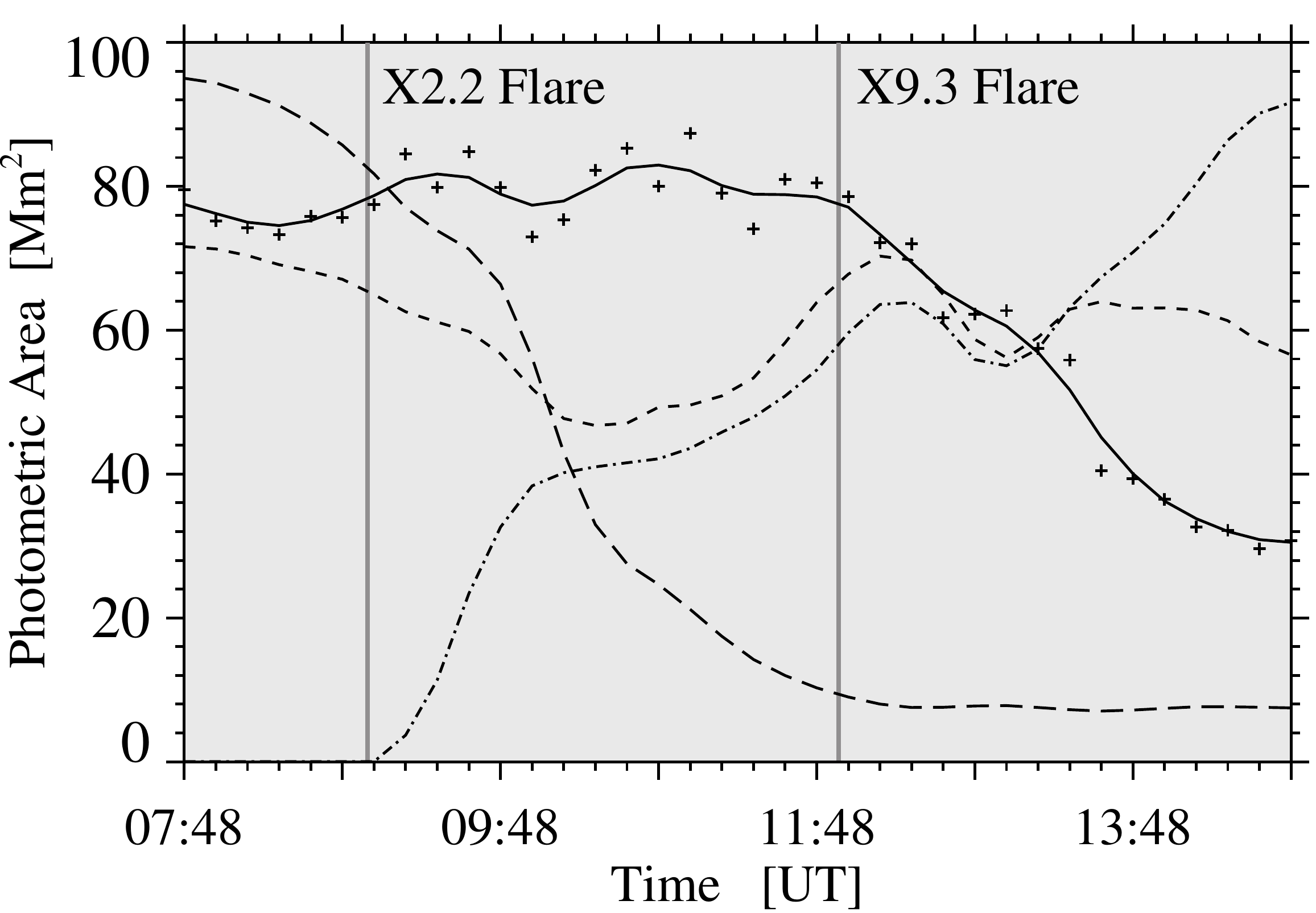}
\hspace*{1mm}
\includegraphics[width=0.74\columnwidth]{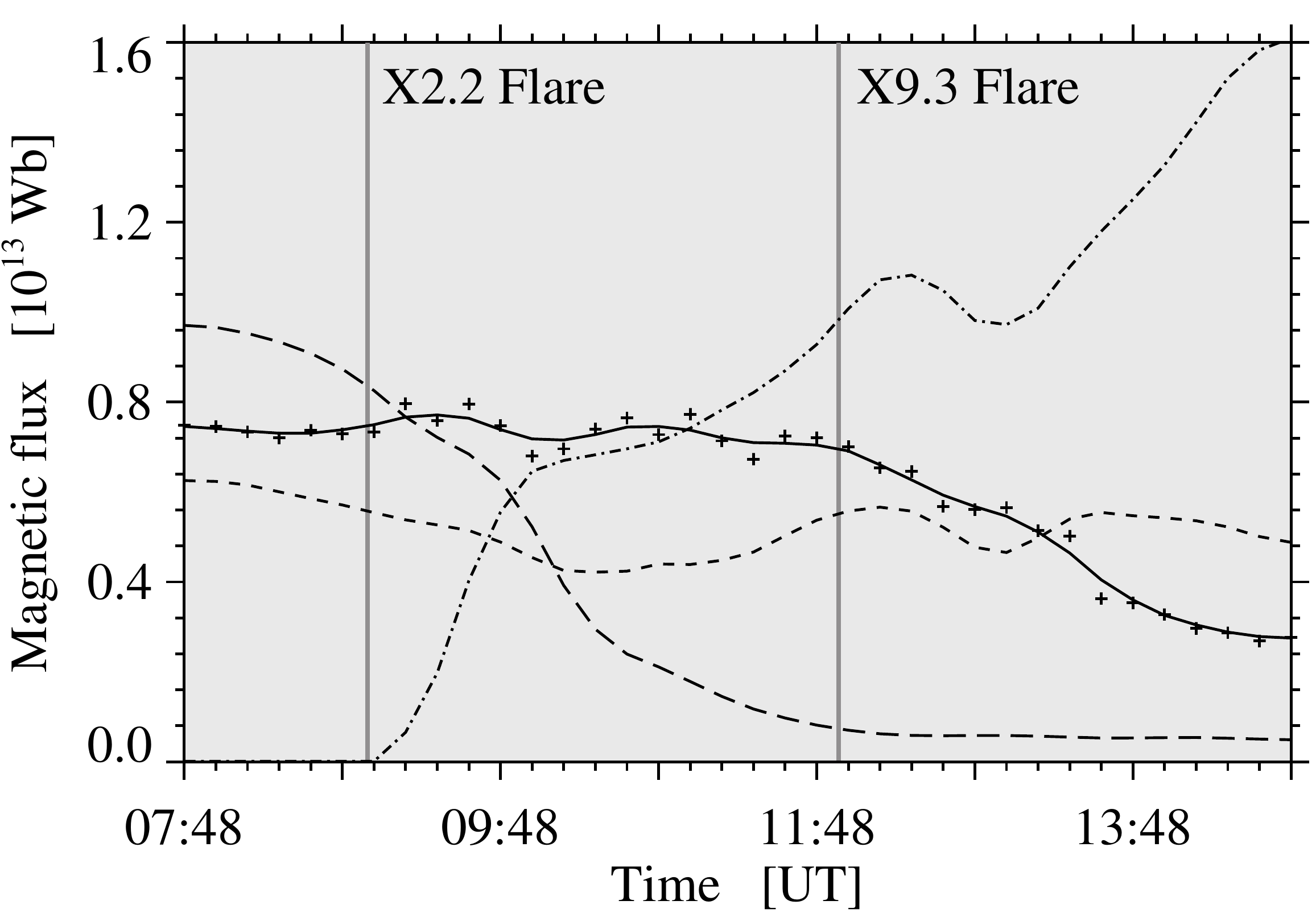}
\hspace*{2mm}
\caption{Evolution of photometric area and magnetic flux for the four regions
\textsf{A} (\textit{solid}), \textsf{B} (\textit{long-dashed}), \textsf{C} 
(\textit{dashed}), and \textsf{D} (\textit{dashed-dotted}) over an 8-hour period
on 6~September 2017. Regions \textsf{A}\,--\,\textsf{C} indicate penumbral decay, 
whereas \textsf{D} is a site of umbral strengthening. The time profiles were
significantly smoothed to avoid clutter. The original data points are shown
as plus signs for region \textsf{A} to give an impression of the scatter.}
\label{FIG05}
\end{SCfigure*}

The regions \textsf{A\,--\,C} and \textsf{D} (as marked in Fig.~\ref{FIG02}) 
were chosen as examples for the regions exhibiting penumbral decay and umbral 
strengthening, respectively (Fig.~\ref{FIG05}). Aligned and geometrical corrected
continuum images and total magnetic flux maps from the inverted vector magnetograms
were used to compute the photometric and magnetic flux evolution in these regions. 
The penumbra and umbra were selected using the intensity threshold of 
$I_\mathrm{pen} < 0.9\,I_{0}$ and $I_\mathrm{umb} < 0.65\,I_{0}$, respectively, 
where $I_{0}$ refers to the normalized quiet-Sun intensity (see column two in 
the online movie). In region \textsf{B}, area and flux decayed after the X2.2 flare, 
whereas these values declined in region \textsf{A} only after the X9.3 flare. 
The penumbra decayed at first in region \textsf{C} after the X2.2 flare but in the 
interim phase, the penumbra replenished and later decayed after the second flare. 
In contrast, area and flux increased in umbral region \textsf{D} after both flares.

To quantify proper motions in and around the $\delta$-spot, we applied
DAVE4VM to vector magnetograms
(see Fig.~\ref{FIG03}). The time-series was divided into the three
aforementioned flare phases. Only some of the flow features were persistent.
Nonetheless, also these region were not static
but continuously exhibited dynamic features in all phases: 
(1) The moat flows around negative polarity spot `d', positive polarity spot 
`b', and the western side of the $\delta$-spot -- the moat flows in sunspot `d' 
and the $\delta$-spot weakened from pre- to post-flare phase. 
(2) The pronounced flow pattern that leads to the separation of the negative 
polarity of the $\delta$-spot and the positive-polarity spot `b' -- the flow vectors 
in northward direction in the $\delta$-spot were well aligned with the PIL. In the 
interim phase, these flows appeared to form an inverted \textsf{S}
tracing the shape of the PIL. These flows had almost identical values in the 
pre- and post-flare phases ($\bar{v}_\mathrm{pre} = 0.49$~km~s$^{-1}$ and 
$\bar{v}_\mathrm{post} = 0.50$~km~s$^{-1}$). The opposing southward flows in the 
southern sunspot became more curved with increased areal coverage from pre-flare 
to interim phase. These flows became somewhat stronger in the interim phase 
($\bar{v}_\mathrm{pre} = 0.35$~km~s$^{-1}$ to $\bar{v}_\mathrm{post} = 
0.37$~km~s$^{-1}$) and traced the circumventing motion of spot `b' around the 
positive polarity of the $\delta$-spot. In general, the velocities along the PIL 
remained high during all three phases with some patches of velocities reaching 
0.5\,--\,0.7~km~s$^{-1}$. 
(3) The velocities in the positive part of the $\delta$-spot remained low. Weak
curling flows exhibited counter-streaming along the upper part of the PIL 
with respect to the northward motion seen in the negative polarity of the 
$\delta$-spot. These flow vectors were seen prominently in the pre-flare 
and interim phases
($\bar{v}_\mathrm{pre} = 0.21$~km~s$^{-1}$ to $\bar{v}_\mathrm{int} = 0.26$~km~s$^{-1}$). 
However, uniform converging motions to the center replaced
the curling motions in the post-flare phase.

The LOS velocity maps based on inversions provided the missing information 
to study flows around the $\delta$-spot in 3D. Only some of the flow features 
were persistent. The Evershed flow appears in the upper/lower sunspots and the 
positive polarity of the $\delta$-spot. Although, only two blueshifted patches 
can be seen in the negative-polarity $\delta$-spot. The flows parallel to the PIL 
displayed successive red/blue/redshifts. This alternating flow pattern remained 
persistent before, between, and after both flares.

\section{Discussion and conclusions}

Peculiar horizontal proper motions are at the center of the present work, which 
were responsible for creating shear and twisting motions powering two X-class 
flares in declining phase of solar cycle 24. \citet{Zirin1993} observed a large 
active region with convoluted magnetic fields along the PIL and with a large curl
of the horizontal magnetic field. The field lines along the PIL reconnected with 
the expanding and stretched-out opposite-polarity regions. Similarly, the extended
and curved PIL of the present $\delta$-spot was the site of many flares, 
specifically two major X-class flares, because of strong magnetic field gradients 
between drawn out umbral cores of opposite polarity in close proximity. 
\citet{Sudol2005} showed that abrupt, significant, and permanent changes of the 
photospheric longitudinal magnetic field are ubiquitous features of X-class flares, 
which were observed for the two X-class flares, too.

Persistent flow patterns along PIL contributed to sheared magnetic field structures:
(1) Strong shear flows affected the negative, parasitic polarity, which was sliding 
along the PIL northward. (2) A group of positive-polarity spots bypassed the main 
spot to the south, and the separation speed increased after the two X-class flares. 
(3) Rapid penumbral decay weakened the moat flow associated with several sunspots 
surrounding the main spot.

\citet{Yang2017} proposed a ``block-induced'' eruption model for the X9.3 flare. 
Newly emerging flux is blocked by already existing flux, which leads to increased
flare-productivity. In addition, they conclude that the X9.3 flare was triggered 
by an erupting filament due to kink instability. However, this flare scenario is 
not new. A similar flare-prolific active region NOAA~5395 was observed by 
\citet{Wang1991}, which exhibited a complex structure, an extended PIL, coalescence 
of spots creating a $\delta$-configuration, expulsion of smaller spots in curved 
trajectories, and strong shear motions. Furthermore, \citet{Denker1998a} noticed 
a similar phenomenon on much smaller scales, i.e., the head-on collision of flux
systems forming a small $\delta$-spot, which produced many C- and M-class flares.
\citet{Tanaka1991} studied a similar $\delta$-spot with complex structure 
and consecutive flares along the PIL. The $\delta$-spot was formed by compressing 
and compacting  opposite polarities. In addition, a light-bridge coincided with 
the PIL with strong horizontal motions. The author proposed that flares in evolving
$\delta$-spots are caused by the emergence of twisted magnetic flux ropes. The 
magnetic topology of such regions is governed by tightly twisted (sheet-like) 
magnetic knots within emerging flux ropes. The elevated flare productivity of such 
regions is likely the result of internal magnetic structure and formation of 
anomalous magnetic ropes.

Active region NOAA~12673 started out as a simple $\alpha$-spot. The emergence 
of multiple bipolar regions to the north and south created a complex active 
region, when first the leading spots of the same polarity as the $\alpha$-spot 
were incorporated in the already existing flux system. The $\delta$-configuration 
was created when the trailing polarity piled up behind the main spot. This 
$\delta$-spot is different than usual $\delta$-configurations because the PIL is 
established within the remnants of the compacted bipolar regions. The parasitic
polarity is surrounded on three sides by the majority polarity, and it is the 
northern side, where the extended and curved umbral cores circumnavigate the 
pre-existing flux system that initiates the flare. This location is also the site
of white-light flare emission in both flares. The observed photospheric shear 
motions created a highly non-potential field configuration, which provided the energy
that powered these, in many aspects homologous, X-class flares 
\citep[e.g.,][]{Harvey1976, Yang2004}. However, differences exist, with the former 
flare being more confined and the later being related to a filament eruption and 
major CME. The emergence of new flux plays a role in the onset of flares. 
\citet{Wang1994} studied five X-class flares and noticed that the magnetic shear
along the PIL increased after all flares. According to authors, the emergence of 
new flux at the onset of each flare had sufficient energy to power flares and 
left the photosphere in a more sheared configuration.

The white-light brightenings associated with two X-class flares are similar
to those in a case study of \citet{Wang2017}, who found low-atmospheric 
small-scale precursors, i.e., brightenings along the PIL before an M6.5 flare. 
These brightenings moved along the PIL instead of away and were confined, while 
the main flare was much more extended, as in the flare model proposed by 
\citet{Kusano2012}. In this model, the pre-flare brightenings can be ascribed to 
heating of the current sheet between the pre-existing large-scale field and 
small-scale intrusions of parasitic polarity forming barb-like structures along 
the PIL. This interaction aided the reconnection of the ambient legs of large-scale
sheared loops rooted in major flux concentration, thus producing precursor brightenings. 
The X2.2 and X9.3 flares may not completely follow this scenario, but the similarity
arises from small-scale brigthenings appearing along the PIL during the confined 
X2.2 flare, followed by more extended flare emission in the X9.3 flare, which was 
associated with a filament eruption and a CME.

\begin{acknowledgements}
SDO HMI and AIA data are provided by the Joint Science Operations Center --
Science Data Processing.
\end{acknowledgements}


\clearpage
\newpage


\begin{appendix}
\onecolumn
\section{Online material}

\noindent\includegraphics[width=\textwidth]{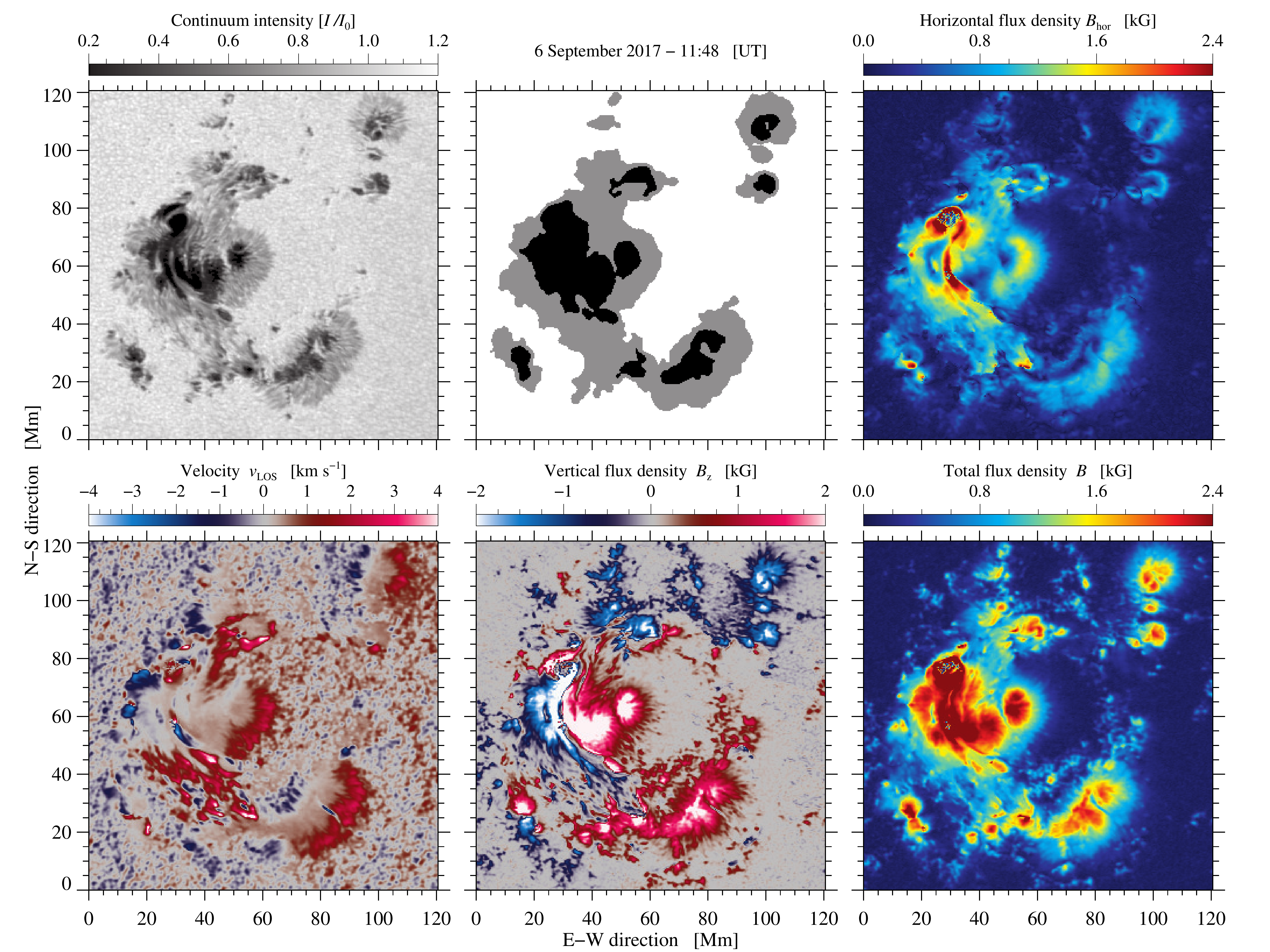}\medskip
\normalsize\footnotesize

\noindent\textbf{Online movie 1:} Results of the spectral inversions 
for the photospheric Fe\,\textsc{i} 6173~\AA\ line. This snapshot shows 
physical maps of active region NOAA~12673 at 11:48~UT on 6~September 2017
just before the X9.3 flare. The whole time-series consists of 35 maps at a
12-minute cadence starting at 07:48~UT: normalized intensity $I / I_0$,
masks of penumbra (gray) and umbra (black), horizontal component of the 
magnetic flux density $B_\mathrm{hor}$, LOS velocity $v_\mathrm{LOS}$, 
vertical component of the magnetic flux density $B_\mathrm{z}$, and total 
flux density (\textit{top-left to bottom-right}).
\normalsize

\end{appendix}


\begin{thebibliography}{42}
\expandafter\ifx\csname natexlab\endcsname\relax\def\natexlab#1{#1}\fi

\bibitem[{{Abramenko} {et~al.}(2010){Abramenko}, {Yurchyshyn}, {Linker},
  {Miki{\'c}}, {Luhmann}, \& {Lee}}]{Abramenko2010}
{Abramenko}, V., {Yurchyshyn}, V., {Linker}, J., {et~al.} 2010, ApJ, 712, 813

\bibitem[{{Bai}(2006)}]{Bai2006}
{Bai}, T. 2006, Sol.\ Phys., 234, 409

\bibitem[{{Beauregard} {et~al.}(2012){Beauregard}, {Verma}, \&
  {Denker}}]{Beauregard2012}
{Beauregard}, L., {Verma}, M., \& {Denker}, C. 2012, AN, 333

\bibitem[{{Bobra} {et~al.}(2014){Bobra}, {Sun}, {Hoeksema}, {Turmon}, {Liu},
  {Hayashi}, {Barnes}, \& {Leka}}]{Bobra2014}
{Bobra}, M.~G., {Sun}, X., {Hoeksema}, J.~T., {et~al.} 2014, Sol.\ Phys., 289,
  3549

\bibitem[{{Borrero} {et~al.}(2011){Borrero}, {Tomczyk}, {Kubo},
  {Socas-Navarro}, {Schou}, {Couvidat}, \& {Bogart}}]{Borrero2011a}
{Borrero}, J.~M., {Tomczyk}, S., {Kubo}, M., {et~al.} 2011, Sol.\ Phys., 273,
  267

\bibitem[{{Brueckner} {et~al.}(1995){Brueckner}, {Howard}, {Koomen},
  {Korendyke}, {Michels}, {Moses}, {Socker}, {Dere}, {Lamy}, {Llebaria},
  {Bout}, {Schwenn}, {Simnett}, {Bedford}, \& {Eyles}}]{Brueckner1995}
{Brueckner}, G.~E., {Howard}, R.~A., {Koomen}, M.~J., {et~al.} 1995, Sol.\
  Phys., 162, 357

\bibitem[{{Deng} {et~al.}(2006){Deng}, {Xu}, {Yang}, {Cao}, {Liu}, {Rimmele},
  {Wang}, \& {Denker}}]{Deng2006}
{Deng}, N., {Xu}, Y., {Yang}, G., {et~al.} 2006, ApJ, 644, 1278

\bibitem[{{Denker} \& {Wang}(1998)}]{Denker1998a}
{Denker}, C. \& {Wang}, H. 1998, ApJ, 502, 493

\bibitem[{{Fritzov{\'a}-{\v S}vestkov{\'a}} \& {{\v
  S}vestka}(1973)}]{Fritzova1973}
{Fritzov{\'a}-{\v S}vestkov{\'a}}, L. \& {{\v S}vestka}, Z. 1973, Sol.\ Phys.,
  29, 417

\bibitem[{{Gopalswamy} {et~al.}(2003){Gopalswamy}, {Lara}, {Yashiro}, {Nunes},
  \& {Howard}}]{Gopalswamy2003b}
{Gopalswamy}, N., {Lara}, A., {Yashiro}, S., {Nunes}, S., \& {Howard}, R.~A.
  2003, in ESA SP, Vol. 535, Solar Variability as an Input to the Earth's
  Environment, ed. A.~{Wilson}, 403--414

\bibitem[{{Harvey} \& {Harvey}(1976)}]{Harvey1976}
{Harvey}, K.~L. \& {Harvey}, J.~W. 1976, Sol.\ Phys., 47, 233

\bibitem[{{Hoeksema} {et~al.}(2014){Hoeksema}, {Liu}, {Hayashi}, {Sun},
  {Schou}, {Couvidat}, {Norton}, {Bobra}, {Centeno}, {Leka}, {Barnes}, \&
  {Turmon}}]{Hoeksema2014}
{Hoeksema}, J.~T., {Liu}, Y., {Hayashi}, K., {et~al.} 2014, Sol.\ Phys., 289,
  3483

\bibitem[{{Kilcik} {et~al.}(2011){Kilcik}, {Yurchyshyn}, {Abramenko}, {Goode},
  {Gopalswamy}, {Ozguc}, \& {Rozelot}}]{Kilcik2011}
{Kilcik}, A., {Yurchyshyn}, V.~B., {Abramenko}, V., {et~al.} 2011, ApJ, 727, 44

\bibitem[{{Kusano} {et~al.}(2012){Kusano}, {Bamba}, {Yamamoto}, {Iida},
  {Toriumi}, \& {Asai}}]{Kusano2012}
{Kusano}, K., {Bamba}, Y., {Yamamoto}, T.~T., {et~al.} 2012, ApJ, 760, 31

\bibitem[{{Lee} {et~al.}(2016){Lee}, {Moon}, \& {Nakariakov}}]{Lee2016}
{Lee}, K., {Moon}, Y.-J., \& {Nakariakov}, V.~M. 2016, ApJ, 831, 131

\bibitem[{{Leka} {et~al.}(2009){Leka}, {Barnes}, {Crouch}, {Metcalf}, {Gary},
  {Jing}, \& {Liu}}]{Leka2009}
{Leka}, K.~D., {Barnes}, G., {Crouch}, A.~D., {et~al.} 2009, Sol.\ Phys., 260,
  83

\bibitem[{{Lemen} {et~al.}(2012){Lemen}, {Title}, {Akin}, {Boerner}, {Chou},
  {Drake}, {Duncan}, {Edwards}, {Friedlaender}, {Heyman}, {Hurlburt}, {Katz},
  {Kushner}, {Levay}, {Lindgren}, {Mathur}, {McFeaters}, {Mitchell}, {Rehse},
  {Schrijver}, {Springer}, {Stern}, {Tarbell}, {Wuelser}, {Wolfson}, {Yanari},
  {Bookbinder}, {Cheimets}, {Caldwell}, {Deluca}, {Gates}, {Golub}, {Park},
  {Podgorski}, {Bush}, {Scherrer}, {Gummin}, {Smith}, {Auker}, {Jerram},
  {Pool}, {Soufli}, {Windt}, {Beardsley}, {Clapp}, {Lang}, \&
  {Waltham}}]{Lemen2012}
{Lemen}, J.~R., {Title}, A.~M., {Akin}, D.~J., {et~al.} 2012, Sol.\ Phys., 275,
  17

\bibitem[{{Liu} {et~al.}(2010){Liu}, {Liu}, {Wang}, {Deng}, \&
  {Wang}}]{Liu2010}
{Liu}, R., {Liu}, C., {Wang}, S., {Deng}, N., \& {Wang}, H. 2010, ApJL, 725,
  L84

\bibitem[{{Metcalf}(1994)}]{Metcalf1994}
{Metcalf}, T.~R. 1994, Sol.\ Phys., 155, 235

\bibitem[{{November} \& {Simon}(1988)}]{November1988}
{November}, L.~J. \& {Simon}, G.~W. 1988, ApJ, 333, 427

\bibitem[{{Pesnell} {et~al.}(2012){Pesnell}, {Thompson}, \&
  {Chamberlin}}]{Pesnell2012}
{Pesnell}, W.~D., {Thompson}, B.~J., \& {Chamberlin}, P.~C. 2012, Sol.\ Phys.,
  275, 3

\bibitem[{{Scherrer} {et~al.}(2012){Scherrer}, {Schou}, {Bush}, {Kosovichev},
  {Bogart}, {Hoeksema}, {Liu}, {Duvall}, {Zhao}, {Title}, {Schrijver},
  {Tarbell}, \& {Tomczyk}}]{Scherrer2012}
{Scherrer}, P.~H., {Schou}, J., {Bush}, R.~I., {et~al.} 2012, Sol.\ Phys., 275,
  207

\bibitem[{{Schuck}(2008)}]{Schuck2008}
{Schuck}, P.~W. 2008, ApJ, 683, 1134

\bibitem[{{Sheeley} \& {Wang}(2015)}]{Sheeley2015}
{Sheeley}, Jr., N.~R. \& {Wang}, Y.-M. 2015, ApJ, 809, 113

\bibitem[{{Sudol} \& {Harvey}(2005)}]{Sudol2005}
{Sudol}, J.~J. \& {Harvey}, J.~W. 2005, ApJ, 635, 647

\bibitem[{{Sun} \& {Norton}(2017)}]{Sun2017}
{Sun}, X. \& {Norton}, A.~A. 2017, Res. Notes AAS, 1, 24

\bibitem[{{Tan} {et~al.}(2009){Tan}, {Chen}, {Abramenko}, \& {Wang}}]{Tan2009}
{Tan}, C., {Chen}, P.~F., {Abramenko}, V., \& {Wang}, H. 2009, ApJ, 690, 1820

\bibitem[{{Tanaka}(1991)}]{Tanaka1991}
{Tanaka}, K. 1991, Sol.\ Phys., 136, 133

\bibitem[{{Temmer} {et~al.}(2003){Temmer}, {Veronig}, \&
  {Hanslmeier}}]{Temmer2003}
{Temmer}, M., {Veronig}, A., \& {Hanslmeier}, A. 2003, Sol.\ Phys., 215, 111

\bibitem[{{{\v S}vestka}(1995)}]{Svestka1995}
{{\v S}vestka}, Z. 1995, Adv.\ Space Res., 16, 27

\bibitem[{{Verma} \& {Denker}(2011)}]{Verma2011}
{Verma}, M. \& {Denker}, C. 2011, A\&A, 529, A153

\bibitem[{{Wang} {et~al.}(1994){Wang}, {Ewell}, {Zirin}, \& {Ai}}]{Wang1994}
{Wang}, H., {Ewell}, Jr., M.~W., {Zirin}, H., \& {Ai}, G. 1994, ApJ, 424, 436

\bibitem[{{Wang} \& {Liu}(2015)}]{Wang2015}
{Wang}, H. \& {Liu}, C. 2015, Res. Astron. \& Astrophys., 15, 145

\bibitem[{{Wang} {et~al.}(2017){Wang}, {Liu}, {Ahn}, {Xu}, {Jing}, {Deng},
  {Huang}, {Liu}, {Kusano}, {Fleishman}, {Gary}, \& {Cao}}]{Wang2017}
{Wang}, H., {Liu}, C., {Ahn}, K., {et~al.} 2017, Nature Astron., 1, 0085

\bibitem[{{Wang} {et~al.}(1991){Wang}, {Tang}, {Zirin}, \& {Ai}}]{Wang1991}
{Wang}, H., {Tang}, F., {Zirin}, H., \& {Ai}, G. 1991, ApJ, 380, 282

\bibitem[{{Wang} {et~al.}(2018){Wang}, {Yurchyshyn}, {Liu}, {Ahn}, {Toriumi},
  \& {Cao}}]{Wang2018}
{Wang}, H., {Yurchyshyn}, V., {Liu}, C., {et~al.} 2018, Res. Notes AAS, 2, 8

\bibitem[{{Wang} {et~al.}(2014){Wang}, {Liu}, {Deng}, \& {Wang}}]{Wang2014}
{Wang}, S., {Liu}, C., {Deng}, N., \& {Wang}, H. 2014, ApJL, 782, L31

\bibitem[{{Welsch} {et~al.}(2009){Welsch}, {Li}, {Schuck}, \&
  {Fisher}}]{Welsch2009}
{Welsch}, B.~T., {Li}, Y., {Schuck}, P.~W., \& {Fisher}, G.~H. 2009, ApJ, 705,
  821

\bibitem[{{Yang} {et~al.}(2004){Yang}, {Xu}, {Cao}, {Wang}, {Denker}, \&
  {Rimmele}}]{Yang2004}
{Yang}, G., {Xu}, Y., {Cao}, W., {et~al.} 2004, ApJL, 617, 151

\bibitem[{{Yang} {et~al.}(2017){Yang}, {Zhang}, {Zhu}, \& {Song}}]{Yang2017}
{Yang}, S., {Zhang}, J., {Zhu}, X., \& {Song}, Q. 2017, ApJL, 849, L21

\bibitem[{{Yurchyshyn} {et~al.}(2006){Yurchyshyn}, {Liu}, {Abramenko}, \&
  {Krall}}]{Yurchyshyn2006}
{Yurchyshyn}, V., {Liu}, C., {Abramenko}, V., \& {Krall}, J. 2006, Sol.\ Phys.,
  239, 317

\bibitem[{{Zirin} \& {Wang}(1993)}]{Zirin1993}
{Zirin}, H. \& {Wang}, H. 1993, Nature, 363, 426

\end{thebibliography}
\end{document}